\documentclass[pre,preprint,showpacs,preprintnumbers,amsmath,amssymb]{revtex4}
\usepackage{graphicx}
\usepackage{dcolumn}
\usepackage{bm}
\usepackage[mathscr]{eucal}
\usepackage{mathrsfs}

\begin{document}

\title{First-principles calculation of the entropy of liquids with a case study on sodium}

\author{Koun Shirai$^{1,2}$, Hiroyoshi Momida$^3$, Kazunori Sato$^4$, and Sangil Hyun$^{5}$}

\affiliation{%
$^{1}$SANKEN, Osaka University,  
8-1 Mihogaoka, Ibaraki, Osaka 567-0047, Japan
}%
\affiliation{%
$^{2}$Vietnam Japan University, Vietnam National University, Hanoi \\
Luu Huu Phuoc Road, My Dinh 1 Ward, Nam Tu Liem District, Hanoi, Vietnam
}%
\affiliation{%
$^{3}$Advanced Materials Laboratory, Sumitomo Electric Industries, Ltd. \\
1-1-1 Koyakita, Itami, Hyogo 664-0016, Japan
}%

\affiliation{$^{4}$Graduate School of Engineering, Osaka University \\
2-1 Yamadaoka, Suita, Osaka 565-0871, Japan}

\affiliation{$^{5}$Korea Institute of Ceramic Engineering and Technology \\
101 Soho-ro, Jinju-si, Gyeongsangnam-do, 52851, Korea
}

\begin{abstract}
Despite increasing demands for the thermodynamic data of liquids in a wide range of science and engineering fields, there is a still a considerable lack of reliable data over a wide range of temperature ($T$) and pressure conditions. The most significant obstacle is that there is no practical method to calculate the entropy ($S$) of liquids. 
This problem can be solved using the thermodynamic definition of entropy, i.e., $S = \int C d\ln T$, where $C$ is specific heat. The specific heat is calculated by the derivative of the internal energy $U$ with respect to $T$. Both quantities, i.e., $U$ and $T$, are well defined in the molecular dynamics (MD) simulations based on density functional theory. 
The reliability of the present method is entirely dependent on the accuracy of the specific heat of liquid, for which there is no standard model. The problem with liquids is that there are no eigenstates, based on which the standard procedures are constructed. The relationship between $U$ and $T$ is affected by the energy relaxation processes, the effect of which appears in the $T$ dependence on the specific heat of liquids. This motivates us to conduct MD simulations by isolating the system from an external heat bath. In this paper, by applying this method to the liquid sodium, it is demonstrated that the experimental $T$ dependence of the isochoric specific heat is reproduced well without any empirical parameter. On this basis, the entropy of the liquid Na is obtained with a good agreement with experimental values.
\end{abstract}

\maketitle

\section{Introduction}
\label{sec:Introduction}
The thermodynamics data of liquids are required in a wide range of scientific and technology fields; however, the lack of reliable free energy data over a desirable range of temperature ($T$) and pressure ($P$) conditions is a serious obstacle impeding the accurate prediction of the thermodynamic behavior of liquids. The area of geothermal fluids is only one such example among many \cite{Stefansson13}. In particular, the thermodynamic functions of liquids near critical points are difficult to predict \cite{Woodcock18}, and knowledge of these functions is required for many important applications  \cite{Fang14}. The presence of the second critical point in metastable water currently has attracted considerable attention \cite{Poole92,Mishima93,Wilding06,Gallo16}; however, the lack of reliable data on the free energy prevents us from reaching a decisive conclusion. In current material research, even under normal conditions, there are many cases where the lack of reliable thermodynamic functions represents a significant disadvantage. It has been demonstrated recently that a diamond can be grown at ambient pressure using a mixed liquid Ga--Fe--Ni--Si \cite{Gong24}. If we could calculate the free energies of the liquid-state reactants, we could predict this reaction successfully. Unfortunately, calculating the thermodynamic functions of multicomponent molten silica is a major challenge in the geophysics domain \cite{Mysen-Richet05}.

The lack of reliable thermodynamic data persists even with the recent development of density functional theory (DFT), which is today acquainted as a standard theory to predict the ground-state properties of solids.
The central difficulty is the calculation of the entropy ($S$) for liquids \cite{Ali19,Holten12}. 
Formally, there is the general formula for the entropy of liquids from the statistical mechanics method. For an $N$-particle system, $S$ is given by
\begin{equation}
S = S_{1}
-  k_{\rm B} \frac{\rho^{N}}{N!} \int g_{N}({\bf r}_{1} \dots {\bf r}_{N}) \ln g_{N}({\bf r}_{1} \dots {\bf r}_{N}) d{\bf r}_{1} \dots d{\bf r}_{N},
\label{eq:S-corr-expansion}
\end{equation}
where $S_{1}=(3/2)-\ln(\rho \Lambda^{3})$ ($\rho$: density and $\Lambda$: the deBroglie wavelength) and $g_{N}({\bf r}_{1} \dots {\bf r}_{N})$ is the $N$-particle correlation function \cite{Wallace94,Baranyai89}. However, in its original form, this formula is not useful due to the intractable form of $g_{N}({\bf r}_{1} \dots {\bf r}_{N})$. Thus, approximation is required to make this formula practically useful. The standard method involves the use of a series expansion of $g_{N}$ with respect to the density $\rho$ \cite{Yokoyama02,Yokoyama02a,Sharma08,Gao18,Widom19}. The $m$-th term $S_{m}$, which contains the $m$-particle correlation function, $g_{N}^{(m)}({\bf r}_{1} \dots {\bf r}_{m})$, is expressed as
\begin{equation}
S_{m} = 
-  k_{\rm B} \frac{\rho^{m}}{m!} \int g_{N}^{(m)}({\bf r}_{1} \dots {\bf r}_{m}) \ln g_{N}^{(m)}({\bf r}_{1} \dots {\bf r}_{m}) d{\bf r}_{1} \dots d{\bf r}_{m}.
\label{eq:S-m}
\end{equation}
In actual applications, it is a common practice to truncate the expansion at the second term $S_{2}$. Based on the recently developed interpretation of the entropy as missing information \cite{Ben-Naim,Ben-Naim18}, some authors attempted to further decompose $S_{2}$ into the fluctuation and mutual information parts \cite{Gao18}.
However, the problem of this method is whether truncating at the second order is sufficient \cite{Gao18,Widom19}. Currently, this method cannot be considered as a universal tool to calculate the entropy of liquids. 

Another approach can be used to calculate entropy. In thermodynamics, entropy is defined by
\begin{equation}
S(T) = \int_{0}^{T} \frac{C(T')}{T'} dT',
\label{eq:thermodynamicS}
\end{equation}
where $C$ is specific heat, provided that the integration is performed along a reversible path. The specific heat is the $T$ derivative of the internal energy $U$, and the internal energy is the time average of the total energy of the system, i.e., $U=\overline{E_{\rm tot}(t)}$, at equilibrium. The total energy is a well-defined quantity in DFT, and $T$ is well described by molecular dynamics (MD) simulation such that $C(T)$ can be calculated exactly using DFT-based MD simulations, which are referred to as first-principles MD (FPMD) simulations. We also refer to this method as {\it the total-energy approach} here.

The authors applied this approach to the problem of the specific heat jump at the glass transition and demonstrated that the observed jump can be explained reasonably \cite{Shirai22-SH,Shirai22-Silica}. The method can be immediately applied to calculate $S$.
Calculating $S$ by Eq.~(\ref{eq:thermodynamicS}) has an advantage in that experimental values of $S$ are obtained just by this method. The total-energy approach is in accordance with the thermodynamic definition of entropy. However, no information is obtainable about the components of entropy. Only the total entropy is the quantity which is observed by the calorimetric method. In this respect, the total-energy approach is also consistent with the thermodynamic definition of entropy.
Now FPMD simulations are widely used for a variety of material researches, and hence it can be truly the universal tool to calculate entropy.

Despite the apparent advantages of the total-energy approach, this method is utilized infrequently to perform real calculations in material researches. When the authors applied this approach to the problem of the specific heat jump at the glass transition \cite{Shirai22-SH,Shirai22-Silica}, almost nobody had used it to calculate the specific heat of liquids.
An obvious reason for the missing applications is the high cost of the FPMD simulations. Computer simulation of a liquid requires a large cell size and a long simulation time. In addition, the integration in Eq.~(\ref{eq:thermodynamicS}) requires a temperature scanning over the entire range from 0 K to a sufficiently high $T$ in the liquid phase. In particular, the accuracy near the melting temperature, $T_{m}$, is crucial because many thermodynamics quantities undergo large jumps, and actual MD simulations exhibit a width of transition, $\Delta T_{m}$, due to the finite-size of $N$ \cite{Shirai22-SH,Shirai22-Silica}. This problem and other error sources have been investigated in a recent study \cite{Shirai-MeltingT}.

A less obvious reason that hinders the application of Eq.~(\ref{eq:thermodynamicS}) to liquids may be a conceptual difficulty in our understanding of the thermodynamic nature of liquids. 
For a long time, there has been no standard theory of the specific heat for liquids, whereas the specific heat is well understood for solids and gases. Recently, several authors have argued that the phonon model can be applied to liquids too \cite{Wallace98, Granato02,Bolmatov12,Trachenko16, Proctor20,Baggioli21}. 
The main idea of them is that the atom motions of a liquid can be considered a collection of vibrations in the local scale, the idea of which was dated back to the defect model of liquids proposed by Frenkel \cite{Frenkel46}. Thus, the phonon energy, $E_{\rm ph}$, can be also defined for a liquid analogously to the crystal case. Then, the internal energy of the system comprises $E_{\rm ph}$ and the correction term $E_{\rm corr}$ as
\begin{equation}
U = E_{\rm ph} + E_{\rm corr}.
\label{eq:model-Etot}
\end{equation}
Note that the substance of $E_{\rm corr}$ depends on the model: for example, the energy of diffusion motions ($E_{d}$) in the model by Bolmatov {\it et al.}~\cite{Bolmatov12,Trachenko16} and the boundary contribution ($E_{b}$) by Wallace ($E_{b}$ is negative in this case) \cite{Wallace98}, in addition to the anharmonic contribution. The critical issue is how many correction terms exist. It is unclear whether the total energy can be decomposed into Eq.~(\ref{eq:model-Etot}) in an unambiguous manner. We will examine the suitability of this decomposition in Sec.~\ref{sec:liquid-result}.

As discussed throughout this paper, the thermodynamic nature of a liquid is that the relationship between $U$ and $T$ is controlled via atom relaxation (or energy dissipation) processes rather than the eigenstates of particles. In fact, there is no concept of the eigenstates of a single particle (or quasi-particle) for liquids. The atom relaxation processes are described automatically by the MD simulation. However, an important element of our approach is the use of the adiabatic simulation, meaning that any thermostat is excluded. Once a heat bath is introduced to control the temperature, energy exchanges occur between the heat bath and the material under investigation, which vitiates the relationship between $U$ and $T$, which is an intrinsic property of a material. The effect of atom relaxation is significant for the glass transition, because the supercooled liquids of glass-forming materials have large viscosities. This is the reason why the total-energy approach was employed to resolve the specific heat jump for glass materials \cite{Shirai22-SH,Shirai22-Silica}. For normal liquids, the advantage of this approach may not be so clear; however, in this paper, we demonstrate that the effect of atom relaxation appears in the temperature dependence of the specific heat for liquids.

The final goal of this study is the calculation of entropy; however, most of the arguments here are devoted to assessing the specific heat value because the reliability of the thermodynamic method based on Eq.~(\ref{eq:thermodynamicS}) is entirely dependent on the accuracy of $C$. This is even a merit of the thermodynamic method. By the abstract nature of entropy, in the statistical approach, there is no point evaluating the correctness of the result during calculation. In contrast, in the thermodynamic method, the calorimetric data for $C$ are easily available, and we have reliable models for $C$ both at low-temperature and high-temperature limits.
In this paper, liquid sodium is examined as a concrete example of the entropy calculation. A large volume of experimental data is available for liquid Na due to its technological importance in nuclear reactor applications.

The remainder of this paper is organized as follows.
Section \ref{sec:theory} outlines the total-energy approach for specific heat. This part has been described in \cite{Shirai22-SH,Shirai22-Silica}, but is again presented here to further analyze the specific heat of liquids. Prior to discussing the simulation results, experimental data for Na are reviewed in Sec.~\ref{sec:Experiment}. This information is presented because we are here mainly concerned with the isochoric specific heat $C_{V}$, while many experimental data are the isobaric specific heat $C_{P}$.
The results of FPMD simulations are presented in Sec.~\ref{sec:Results}, by analyzing the solid and liquid states separately. Section \ref{sec:Conclusion} summarizes the present results.




\section{Theory of specific heat}
\label{sec:theory}

\subsection{Total-energy approach}
\label{sec:Sim-melt}

\paragraph{General formulation.}
Because DFT is a ground-state theory, it may be necessary to explain the applicability of DFT to the liquid state, which is not the ground state of a material. With the exception of liquid helium, liquids exist only at finite temperatures. An extension of DFT to finite-temperature systems was formulated previously by Mermin, who established that the entropy of electrons is given by a functional of the equilibrium electron density \cite{Mermin65}. Recently, this notion of the electronic entropy was recast within the general framework of the maximum-entropy principle by Yousefi and Caticha \cite{Yousefi24}. Thus, in principle, thermodynamic functions of a liquid can be expressed as a functional of the equilibrium electron density at a given temperature, $\bar{\rho}({\bf r}, T)$. Here, the obvious problem is that the concrete functional form is difficult to know.
However, knowing the concrete form of the functional appears unnecessary. When the Born-Oppenheimer approximation is assumed, atom motions even at high temperatures can be considered for the electron system as the ground state for the instantaneous atom positions. The entropy of a system can be obtained by Eq.~(\ref{eq:thermodynamicS}) through MD simulations, as explained below.

Here, we consider an MD simulation based on DFT. Assume that the system under consideration is composed of $N$ atoms, irrespective of whether the system is a solid or a liquid. The $j$th atom having mass $M_{j}$ is at position ${\bf R}_{j}(t)$ with velocity ${\bf v}_{j}(t)$ at time $t$. In the Born-Oppenheimer approximation, the total energy, $E_{\rm tot}$, of the system is given by the sum of the kinetic energy, $E_{\rm K}$, and the potential energy, $E_{\rm P}$, as
\begin{equation}
E_{\rm tot}(t) \equiv  E_{\rm P} + E_{\rm K} = 
    E_{\rm gs}( \{ {\bf R}_{j}(t) \}) + \frac{1}{2} \sum_{j} M_{j} v_{j}(t)^{2},
\label{eq:total-energy}
\end{equation}
where $E_{\rm gs}( \{ {\bf R}_{j}(t) \})$ is the ground state energy of the electrons, including the ion-ion potentials, for the instantaneous positions $\{ {\bf R}_{j}(t) \}$.
The internal energy in thermodynamics is defined at equilibrium and is given by the time average of the microscopic total energy $E_{\rm tot}(t)$ of the system as
\begin{equation}
U = \overline{E_{\rm tot}(t)} = \overline{E_{\rm gs}( \{ {\bf R}_{j}(t) \})}
    + \frac{1}{2} \sum_{j} M_{j} \overline{ v_{j}(t)^{2} }.
\label{eq:internal-energy}
\end{equation}
When the volume, $V$, of the system is fixed, the isochoric specific heat, $C_{V}$, is obtained by $C_{V} = (\partial U/\partial T)_{V}$. The isobaric specific heat, $C_{P}$, is obtained by adding the contribution of the thermal expansion, $C_{\rm te}$, as
\begin{equation}
C_{P} = C_{V} + C_{\rm te}.
\label{eq:Cp}
\end{equation}
$C_{\rm te}$ is calculated by
\begin{equation}
C_{\rm te} = TV \frac{(\alpha_{P})^{2}}{\kappa_{T}} ,
\label{eq:Cte}
\end{equation}
where $\alpha_{P}$ is the isobaric thermal expansion coefficient and $\kappa_{T}$ is the isothermal compressibility. It is useful to know that, in the limiting case of ideal gases, $\kappa_{T}$ becomes $\kappa_{T} = 1/P$ and $\alpha_{P}$ becomes $\alpha_{P} = 1/T$, which leads to the well-know formula, $C_{\rm te}=R$ ($R$ is the gas constant). For solids, the contribution of the thermal expansion is normally small, i.e., $C_{\rm te} \ll R$.

Here, the electronic contribution to the specific heat ($C_{\rm el}$) and thereby entropy ($S_{\rm el}$) is ignored. In fact, in metallic liquids, it is known that $S_{\rm el}$ is only a tiny fraction of the total entropy \cite{Wallace97a,Wallace98}. However, in Supplemental material, the effect of the Fermi-Dirac distribution at finite temperatures is examined, concluding that the ignorance of $S_{\rm el}$ is rational.

\paragraph{Solid case.}
For solids, the constituent atoms fluctuate around their equilibrium positions. The instantaneous position of the $j$th atom can be expressed by the sum of the equilibrium position, $\bar{\bf R}_{j}$, and a small displacement from this position, $\bar{\bf u}_{j}$, as ${\bf R}_{j}(t) = \bar{\bf R}_{j} + {\bf u}_{j}(t)$. The ground-state energy, $E_{\rm gs}( \{ {\bf R}_{j}(t) \} )$, can be expanded in the Taylor series in terms of this displacement as
\begin{equation}
E_{\rm gs}( \{ {\bf R}_{j}(t) \} ) = 
E_{\rm st}( \{ \bar{\bf R}_{j} \} )
+ \frac{1}{2} \sum_{i,j}  {\bf u}_{i}(t) \cdot {\bf D}_{ij} \cdot {\bf u}_{j}(t).
\label{eq:gs-Taylor}
\end{equation}
The second term on the right-hand side of Eq.~(\ref{eq:gs-Taylor}) is denoted by $E_{\rm P, vib}$.
This part together with $E_{\rm K}$ constitutes the phonon energy and its time average $E_{\rm ph}$ is expressed as
\begin{equation}
E_{\rm ph} = \frac{1}{2} \sum_{i,j} \overline{ {\bf u}_{i}(t)\cdot {\bf D}_{ij} \cdot {\bf u}_{j}(t)}
    + \frac{1}{2} \sum_{j} M_{j} \overline{ v_{j}(t)^{2} },
\label{eq:phonon-energy}
\end{equation}
where ${\bf D}_{ij}$ is the dynamic matrix between the $i$th and $j$th atoms \cite{BornHuang}. 
The remaining part, $E_{\rm st}( \{ \bar{\bf R}_{j} \} )$, in the expansion (\ref{eq:gs-Taylor}) is constant with respect to time and is referred to as the {\em structural} energy. Thus, the internal energy $U_{V}$ is decomposed of
\begin{equation}
U_{V} \equiv U(T,\{ \bar{\bf R}_{j} \} )_{V} = E_{\rm st}( \{ \bar{\bf R}_{j} \} )  +  E_{\rm ph}(T),
\label{eq:U-solid}
\end{equation}
when $V$ is fixed.
Correspondingly, the isochoric specific heat $C_{V}$ is given by the sum of their respective derivatives,
\begin{equation}
C_{V} = C_{\rm st}  +  C_{\rm ph}.
\label{eq:Cv-solid}
\end{equation}
At low temperatures, at which the harmonic approximation holds well, there is no temperature dependence on $E_{\rm st}$, and accordingly $C_{V}$ is determined by $C_{\rm ph}$ alone. At high temperatures, $E_{\rm st}$ comes to appear in $C_{V}$. Note that $C_{\rm st}$ contains all the anharmonic effects, such as frequency shift and phonon-phonon interactions \cite{Leibfried61,Cowely64}.
When $V$ is changed, the thermal expansion energy, $E_{\rm te}(V)$, which leads to Eq.~(\ref{eq:Cte}), should be added to obtain the total internal energy $U$. 

By diagonalizing the dynamic matrix, $E_{\rm ph}$ can be obtained as
\begin{equation}
E_{\rm ph} = \int \hbar \omega \left( \bar{n}(\omega) + \frac{1}{2} \right) g(\omega) d\omega,
\label{eq:phonon-energy-integral}
\end{equation}
where $\hbar$ is the Planck constant, $\bar{n}(\omega)$ is the Bose occupation number, and $g(\omega)$ is the phonon DOS. 
The phonon contribution to the specific heat can be obtained by the analytical form,
\begin{equation}
C_{\rm ph}(T) =
k_{\rm B} \int \left( \frac{\hbar \omega}{k_{\rm B} T} \right)^{2} 
\frac{e^{\hbar \omega/k_{\rm B}T}}{(e^{\hbar \omega/k_{\rm B}T}-1)^{2}}
g(\omega) d\omega,
\label{eq:C-phonon}
\end{equation}
as a function of $T$, where $k_{\rm B}$ is Boltzmann's constant. In the MD simulations, the phonon DOS is obtained from the frequency spectrum, $f_{v}(\omega)$, of atom velocity $v(t)$,
\begin{equation}
g(\omega) = 
 \frac{1}{3 N \frac{3}{2} k_{\rm B}T}  \sum_{j}^{3N} \frac{1}{2} M_{j} f_{v_{j}}^{\ast}(\omega) f_{v_{j}}(\omega).
\label{eq:phononDOS}
\end{equation}
Here, the subscript $j$ denotes a composite index referring to the atom index and the Cartesian coordinates. The conservation of the number of freedoms can be checked by
\begin{equation}
\int g(\omega) d\omega = 1.
\label{eq:sum-DOS}
\end{equation}
Because atom motions are treated classically in conventional MD simulations, the kinetic energy of atoms in the simulations does not obey the Bose-Einstein statistics. Instead, it turns to obey the classical equipartition law of energy,
\begin{equation}
E_{\rm P,vib} = E_{\rm K} = \frac{3}{2} N k_{\rm B}T.
\label{eq:equi-partition}
\end{equation}
To fix this problem, first, this classical term in $\overline{E_{\rm gs}(t)}$ is removed from the total energy, giving the structural energy $E_{\rm st}$,
\begin{equation}
E_{\rm st} = \overline{E_{\rm gs}(t)}-E_{\rm K} = \overline{E_{\rm gs}(t)} - \frac{1}{2} E_{\rm ph}.
\label{eq:Est}
\end{equation}
Then, the phonon energy $E_{\rm ph}$ is evaluated using Eq.~(\ref{eq:phonon-energy-integral}). 
Finally, $E_{\rm tot}$ is recalculated by adding this $E_{\rm ph}$ to $E_{\rm st}$, thereby recovering the Bose-Einstein statistics. Practically, this modification of $E_{\rm tot}$ is important only at very low temperatures. 

\paragraph{Liquid case.}
For liquids, the atoms do not have unique equilibrium positions; thus the expansion (\ref{eq:phonon-energy}) does not make sense. We should stop at the general formula for the internal energy, Eq.~(\ref{eq:internal-energy}), which is valid for liquids too. By taking numerical derivative of $U(T)$ with respect to $T$, we can obtain $C_{V}$. 
Nonetheless, recently, an increasing number of authors adapted the phonon model to analyze the specific heat of liquids \cite{Wallace98, Bolmatov12,Trachenko16, Proctor20,Baggioli21}, because phonon-like DOS can be obtained by applying Eq.~(\ref{eq:phononDOS}). Deriving the phonon dispersion in this manner is  practiced in inelastic neutron scattering analysis \cite{Egelstaff-2ed,Copey74,Smith17}. Hence, it is of interest to investigate how good the phonon model works for liquids in FPMD simulations.

When applying Eq.~(\ref{eq:phononDOS}) to a liquid, care is needed in treating zero-frequency modes. When a solid melts, the atoms begin to have diffusional motions and nonvanishing zero-frequency components $v_{j}(\omega=0)$ appear,
\begin{equation}
v_{j}(\omega=0) = \frac{1}{t_{\rm sm}} \int v_{j}(t) dt,
\label{eq:zero-freq-v}
\end{equation}
where $t_{\rm sm}$ is the simulation time. 
Let $x$ denote the fraction of the zero-frequency modes in the phonon DOS, and thus
\begin{equation}
\int_{\omega>0} g(\omega) d\omega = 1-x.
\label{eq:zeroFreq-modes}
\end{equation}
This zero-frequency component creates the kinetic energy of pure translational motion, $E_{\rm tr} = \sum_{j} (1/2) M_{j} v_{j}(\omega=0)^{2}$, while there is no corresponding potential energy.
Thus, the total energy with constant $V$ comprises three components as
\begin{equation}
U_{V} = E_{\rm st}( \{ \bar{\bf R}_{j} \} )  +  E_{\rm tr}(T) + E_{\rm ph}(T).
\label{eq:internal-energy-4}
\end{equation}
The specific heat component corresponding to $E_{\rm tr}$ is denoted $C_{\rm tr}$. Under the assumption that the phonon model is valid for liquids, the fraction of the zero-frequency modes, $x$, controls the $T$ dependence of $C_{V}$.
When $T$ is such a high temperature, at which all phonons are thermally activated, then the equipartition law of energy holds, and classical limit becomes:
\begin{equation}
C_{\rm tr} + C_{\rm ph} = \left[ \frac{3}{2} x +  \frac{6}{2} (1-x) \right] R
\label{eq:Cliquid-expect}
\end{equation}
This must explain the $T$ dependence of $C_{V}$ of liquids. 
When Wallace compiled the experimental data of many metallic liquids, he observed the general behavior of $C_{V}$ of these liquids that $C_{V}$ is close to $3R$ just above $T_{m}$ and is gradually reduced to approximately $2R$ at the boiling temperature $T_{b}$ \cite{Wallace98,Wallace02}.
At $T=T_{m}$, the diffusional motions begin to occur; thus $x \approx 0$. The longitudinal mode exists over the entire $T$ range as the sound wave. Wallace pointed out that the transverse modes also exist for liquids because the shear viscosity plays a role in restoring the force for the transverse displacements. Therefore, it is reasonable to observe $C_{V} \approx 3R$ just above $T_{m}$. However, as $T$ increases, the viscosity is weakened, and the contribution of the transverse modes gradually diminishes, leaving the longitudinal mode as the solely active mode. Equation (\ref{eq:Cliquid-expect}) predicts $C_{V} = 2R$, when $x=2/3$. The validity of this model is tested in Sec.~\ref{sec:liquid-result}.

\subsection{Equilibrium nature of liquid}
\label{sec:Eq-liquids}
The phonon model is appealing in that the substance giving rise to the specific heat is easy to imagine and analytic formulae are available. In contrast, in the total-energy approach, there is no visible substance giving the specific heat. The only explanation is that the total energy merely gives this value. However, accountability and accuracy differ. Here, the fundamental problem of the phonon model is discussed on the basis of statistical mechanics.

When the partition function of a $N$-particle system,
\begin{equation}
Z = \sum_{r} \exp( -\beta E_{r}),
\label{eq:partition-f}
\end{equation}
is investigated in textbooks of statistical mechanics, we tacitly assume that the system has a set of eigenstates $\{ r \}$, where $\beta=1/k_{\rm B}T$. When the system comprises independent particles (or quasi-particles), the total energy of the system, $E_{r}$, is given by the sum of the eigen-energies, $\epsilon_{s}$, of individual particles,
\begin{equation}
E_{r} = \sum_{s} n_{s} \epsilon_{s}.
\label{eq:Er=sum-s}
\end{equation}
In quantum mechanics, the eigenstate means an infinite lifetime. The equilibrium distribution $\bar{n}_{s}$ is determined by the Bose-Einstein statistics for phonon systems, which is derived on the assumption of independent particles (\cite{Reif}, Sec.~9.2). The total excitation energy of the phonon system is the sum of the phonon energies of the individual phonon modes, as in Eq.~(\ref{eq:phonon-energy-integral}).
In DFT, it is well known that the total energy is not the sum of the energies of the individual atoms. The additive property of Eq.~(\ref{eq:Er=sum-s}) results from the harmonic approximation for the atom excitations. The phonon modes diagonalize the Hamiltonian of the system, and thereby the interactions between different phonons can be omitted, from which the independent-particle description holds for the phonon system. In this manner, the lifetime of a phonon can be regarded infinite.

The problem with liquids is that there are no eigenstates in the sense that there are no modes that last for a long period. The eigenstates are orthogonal to each other and thus there is no overlap between them, which guarantees the independent-particle description. In the literature \cite{Wallace98, Bolmatov12,Trachenko16, Proctor20,Baggioli21}, however, by observing locally oscillatory behavior in atom motions or instantaneous motions, the phonon model is often applied to liquids. This idea is promoted by the practical utility of the vibrational spectrum based on Eq.~(\ref{eq:phononDOS}). The mathematical form of this equation is liable to hide the independent-particle assumption on which the phonon statistics are based.
Phonon-like excitations in a liquid are destroyed soon after being created and are converting to other modes. Their lifetimes are too short to qualify as eigenstates. The phonon-like excitations of a liquid do not diagonalize the Hamiltonian. Thus, they overlap each other, and consequently the summing property, Eq.~(\ref{eq:Er=sum-s}), does not hold. Equation (\ref{eq:phonon-energy-integral}) cannot be applied to liquids. A consequence of the violation of the independent-particle assumption for a liquid is illustrated in Sec.~\ref{sec:liquid-result}. 

Let us describe how the equilibrium is achieved when the system has no eigenstates.
In the MD simulations of a liquid, there is an instantaneous state $i$ with its total energy $E_{i}$. By collecting $i$ over simulations, we can define the probability (population) $P_{i}$ of the instantaneous state $i$. Here, indexes $r$ and $s$ are reserved for presenting eigenstates, while $i$ and $j$ are used to present instantaneous state. There is no guarantee that $P_{i}$ obeys the Bose-Einstein statistics, as explained above. Instead, $P_{i}$ is determined by the balance of transitions,
\begin{equation}
\frac{d}{dt} P_{i} = \sum_{j} W_{ji} P_{j} - \sum_{j} W_{ij} P_{j} ,
\label{eq:MasterEq}
\end{equation}
where $W_{ji}$ is the time-dependent transition probability from state $j$ to $i$. Equation (\ref{eq:MasterEq}) is known as the master equation for stochastic processes \cite{Reif}. The time-dependent transition probability $W_{ji}(t)$ is expressed as $W_{ji}(t) = \nu_{0} \exp( -\gamma_{ji} t)$, where $\gamma_{ji}$ is the transition rate of state $j$ to $i$ and $\nu_{0}$ is the attempt frequency. 
Here, $\gamma_{ji}$ is determined by the energy barrier $E_{b,ji}$ from state $j$ toward $i$ as $\gamma_{ji} = E_{b,ji}/\hbar$ \cite{Eyring64}. The equilibrium of the system is established by the detailed balance $(d/dt) P_{i}=0$ in Eq.~(\ref{eq:MasterEq}). When the states $i$ and $j$ are eigenstates $r$ and $s$ with infinite lifetimes, it has a sense to consider the limit of $t \rightarrow \infty$, namely, 
\begin{equation}
t \gg \tau_{rs},
\label{eq:Equilibrin-cond1}
\end{equation}
for any transition between $r$ and $s$, where $\tau_{rs} = 1/\gamma_{rs}$. 
Under this condition, the transition probability becomes a time-independent constant, 
\begin{equation}
W_{rs} = \nu_{0} \exp(-\beta E_{b,rs}).
\label{eq:Wrs-barrier}
\end{equation}
Then, the population ratio $P_{s}$ to $P_{r}$ is determined by the ratio of the forward and backward transition rates
\begin{equation}
\frac{P_{s}}{P_{r}} =  \frac{W_{rs}}{W_{sr}} = \exp( -\beta (E_{b,rs} - E_{b,sr})) = \exp( -\beta (E_{r} - E_{s})).
\label{eq:detailed-balance}
\end{equation}
In the last step in Eq.~(\ref{eq:detailed-balance}), the barrier height between $s$ and $r$ is dropped by taking the difference between the forward and backward transitions; thus, as expected, the ratio $P_{s}/P_{r}$ is determined by only the energy difference $E_{r} - E_{s}$. 

However, for liquids, we cannot ignore the finite relaxation time; thus, the condition (\ref{eq:Equilibrin-cond1}) is no longer valid. The equilibrium of a liquid is controlled by the dynamic balance between the incessant creation and destruction of phonon-like excitations. The balance $(d/dt) P_{i}=0$ holds only as the time average,
\begin{equation}
\langle \nu_{0} \exp( -\gamma_{ji} t) P_{j} \rangle = \langle \nu_{0} \exp( -\gamma_{ji} t) P_{j} \rangle.
\label{eq:dynamic-balance}
\end{equation}
The brackets designate the time average.
Accordingly, the population $P_{i}$ is determined by the energy barriers in addition to the energy difference between the potential minima \cite{note-effectofbarrier}. $P_{i}$ has dependence on temperature, differently from the phonon systems, where the phonon DOS, $g(\omega)$, is independent of temperature. Because of the importance of the energy relaxation processes, the use of adiabatic MD simulations is essential. Introducing a heat bath alters the energy relaxation processes, which destroys the intrinsic relationship between $U$ and $T$. For the same reason, the pressure bath should also be avoided. Thus, our method employs an $NVE$ ensemble. Consequently, we are mainly concerned with $C_{V}$. The part of $C_{\rm te}$ can be, in principle, calculated by FPMD simulations, through evaluating Eq.~(\ref{eq:Cte}) with additional calculations for $\alpha_{P}(T)$ and $\kappa_{T}(T)$. However, for the present example of Na, the behavior of $C_{\rm te}$ is well understood experimentally, as described in Sec.~\ref{sec:Experiment}, and therefore we utilize the experimental value for $C_{\rm te}$.

\subsection{Calculation conditions}
\label{sec:Calc-cond}
The present FPMD simulations were performed using the Phase/0 code \cite{PHASE}, which employs a pseudopotential method using planewave expansion.
The cutoff energy of 18 Ry was used for the planewave expansion. GGA and norm-conserved pseudopotential are used. Several conditions were tested for the convergence, including the $k$ mesh of the Brilloiun zone, the size of supercell, the effect of smearing of electronic occupancies. These concernings are described in Supplemental material.
The MD simulations employed a time step of 1.2 fs and the total simulation time, $t_{\rm sm}$, was 2.4 ps, depending on the relaxation time. The volume of the crystal, $V$, was fixed at the experimental value.



\section{Review of experiments}
\label{sec:Experiment}
As discussed previously, we focus on the isochoric specific heat $C_{V}$ as the primary quantity to be calculated by the FPMD simulation. In most experiments, however, the obtained specific heat is the isobaric specific heat $C_{P}$. These two specific heats are related through the thermal expansion contribution $C_{\rm te}$, as in Eq.~(\ref{eq:Cp}). In principle, this part $C_{\rm te}$ can also be calculated by FPMD simulations too. However, here, we use the experimental value of $C_{\rm te}$ to obtain $C_{P}$ due to our interest in $C_{V}$. In the following, we review experimental data on the specific heat of Na in order to better compare it to the experimental results. 

\begin{figure}[ht!]
\centering
    \includegraphics[width=120mm, bb=0 0 440 280]{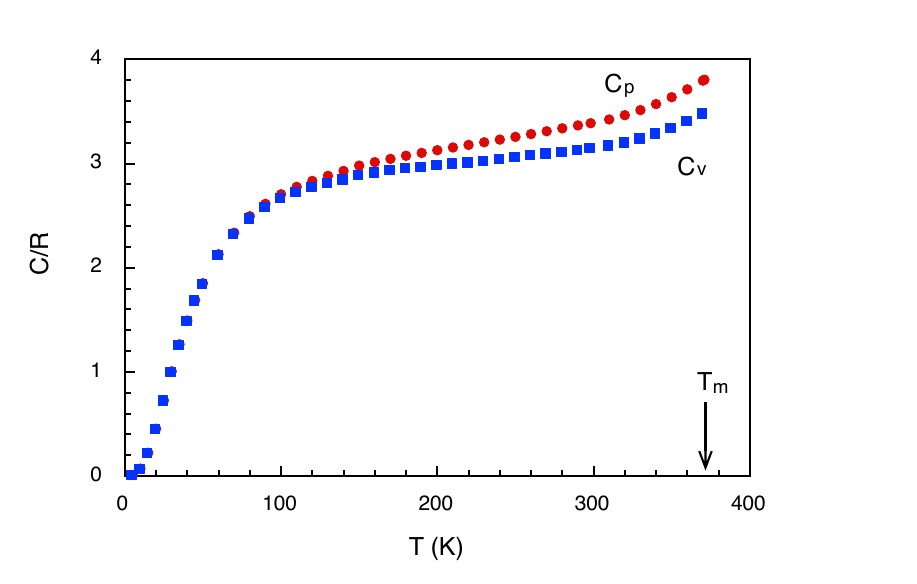} 
\caption{
Experimental values of $C_{P}$ \cite{Alcock94} and the derived values of $C_{V}$ for the solid Na. The melting temperature, $T_{m}$, is marked by the arrow.
}
\label{fig:Alcock}
\end{figure}

A large amount of experimental data on the specific heat of solid Na is available. Only a few are cited here \cite{Ginnings50,Martin60,Martin67}. For the solid Na, there seems no serious problem to standardize $C_{P}(T)$ data. By reviewing the experimental data on $C_{P}(T)$ in various ranges of $T$, Alcock {\it et al.}~created the recommended curve $C_{P}(T)$ \cite{Alcock94}. The data of $C_{P}$ shown in Fig.~\ref{fig:Alcock} are plots from the recommended values in Tables 3.9 and 3.10 of \cite{Alcock94}. $C_{V}$ can be obtained from $C_{P}$ by subtracting $C_{\rm te}$, as in Eq.~(\ref{eq:Cp}).
In the expression for $C_{\rm te}$, Eq.~(\ref{eq:Cte}), all of the quantities, $V$, $\alpha_{P}$, and $\kappa_{T}$, have $T$ dependence to some extent. However, $V$ and $\kappa_{T}$ do not change so much as $T$ changes; thus, the constant values of $V=22.69 \ {\rm cm^{3}/mol}$ and $\kappa_{T}=0.147 \ {\rm GPa^{-1}}$ are adapted. In contrast, the $T$ dependence on $\alpha_{P}$ is significant: in fact, $\alpha_{P}(T)$ asymptotically vanishes as $T \rightarrow 0$. The $T$ dependent $\alpha_{P}$ was reported by Siegel and Quimby in a $T$ range from 80 to 290 K \cite{Siegel38}. From these data along with its extrapolation outside this range, $C_{V}(T)$ is obtained, as plotted in Fig.~\ref{fig:Alcock}. $C_{V}$ reaches the classical limit, $3R$, at $T=200$ K, which is consistent with the Debye temperature of Na, $\Theta_{D}=157$K \cite{Alcock94}.

\begin{figure}[ht!]
\centering
    \includegraphics[width=120mm, bb=0 0 420 330]{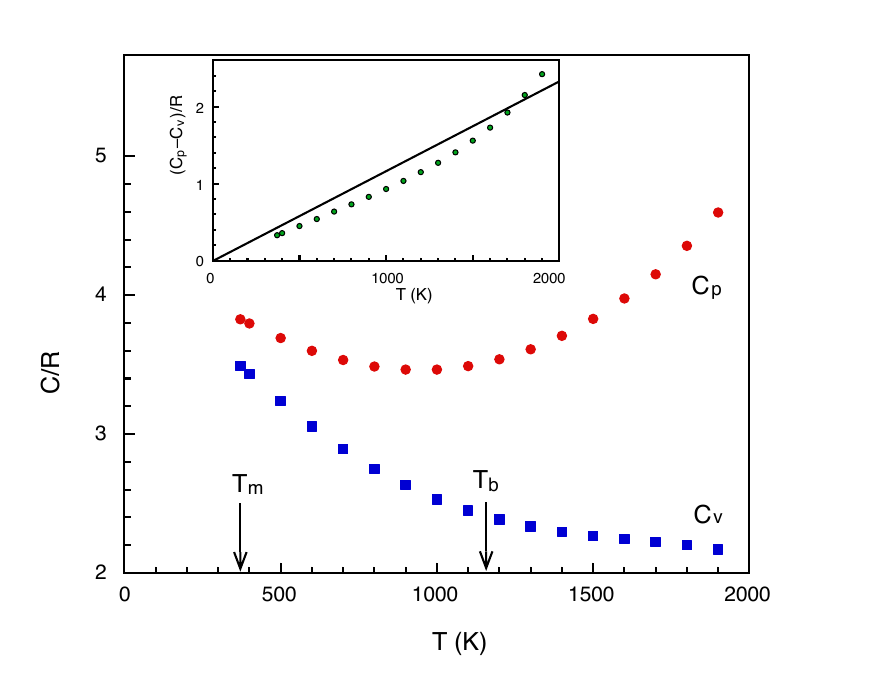} 
\caption{
Experimental values of $C_{P}$ and $C_{V}$ of the liquid Na \cite{Fink95}.
The inset shows the difference $C_{P}-C_{V}$. The line indicates the $C_{\rm te}$ values calculated using Eq.~(\ref{eq:Cte}) with constant $\alpha_{P}$ and $\kappa_{T}$.
}
\label{fig:Fink}
\end{figure}
Care is needed when considering liquids. The fact that the measurement of a liquid is typically performed under the conditions of coexistence of the liquid and the vapor complicates the procedure. Fortunately, a comprehensive review of the liquid Na by Fink and Leibowitz is available \cite{Fink95}. Their recommended values for $C_{P}$ and $C_{V}$ are plotted in Fig.~\ref{fig:Fink}. By compiling the reported values of enthalpy, $H$, of the liquid along the saturation curve, Fink and Leibowitz obtained the specific heat, $C_{\sigma}$, along the saturation curve as
\begin{equation}
C_{\sigma} = \left( \frac{\partial H}{\partial T} \right)_{\sigma} - V \gamma_{\sigma},
\label{eq:Csig}
\end{equation}
where $\gamma_{\sigma} = (\partial P/\partial T)_{\sigma}$. Here, the subscript $\sigma$ denotes that the derivative is taken along the saturation curve. 
Then, they converted it to $C_{P}$ by
\begin{equation}
C_{P} = C_{\sigma} + TV \gamma_{\sigma} \alpha_{P}.
\label{eq:Cp-Csig}
\end{equation}
Finally, $C_{V}$ is obtained using the following thermodynamic relation,
\begin{equation}
C_{V} = C_{P} \frac{\kappa_{S}}{\kappa_{T}},
\label{eq:Cv=Cp-kappa}
\end{equation}
where $\kappa_{S}$ and $\kappa_{T}$ are the adiabatic and isothermal compressibilities, respectively. Thus, the $C_{P}$ values plotted in Fig.~\ref{fig:Fink} are not $C_{P}$ values at 1 atm but those at the saturated vapor pressures. This is why the specific heat of the liquid can be plotted beyond the boiling temperature, $T_{b}=1156$ K. 
Generally, it is known that $\gamma_{\sigma}$ extraordinarily increases with $T$ near the critical point, and accordingly strong $P$ dependence on $C_{P}$ appears. Volatile liquids, such as rare-gas liquids, are representative examples of this case, where the difference $C_{P}-C_{V}$ exceeds $R$, i.e., the value of ideal gas \cite{Gladun66,Gladun71}. The divergent property in $C_{P}$ near the critical point has been debated in \cite{Sengers15,Woodcock17,Woodcock18}. 
For the present case of Na, Fink and Leibowitz pointed out that $\alpha_{P}$ becomes greater than $1/T$, i.e., the value of ideal gas, for $T>2000$K \cite{Fink95}. Because the critical temperature of Na is very high, $T_{c}=2503.7$ K \cite{Fink95}; thus, we can consider the data $C_{P}$ plotted in Fig.~\ref{fig:Fink} as the values at 1 atm for $T<T_{b}$. A recent study demonstrated that the pressure dependence of $C_{P}$ is approximately $0.7 R$/GPa \cite{Li17}, which indicates that its change in $P$ can be ignored around ambient pressure.

As shown in Fig.~\ref{fig:Fink}, $C_{V}$ decreases monotonously as $T$ increases. In contrast, although $C_{P}$ initially decreases above $T_{m}$, it turns to increase above $900$K. The difference $C_{P}-C_{V}$ is shown in the inset of Fig.~\ref{fig:Fink}, where a linear approximation of Eq.~(\ref{eq:Cte}) is indicated by the line. This line was obtained using the constant values of $V=24.84 \ {\rm cm^{3}/mol}$, $\kappa_{T}=0.168 \ {\rm GPa^{-1}}$, and $\alpha_{P}=2.57 \times 10^{-4}\ K^{-1}$ \cite{Wallace97b}, and is consequently expressed as
\begin{equation}
C_{\rm te} = 1.17 \times 10^{-3} RT.
\label{eq:Cte-liquid}
\end{equation}
The good fitting of the linear extrapolation to the experimental data $C_{P}-C_{V}$ confirms that the use of Eq.~(\ref{eq:Cte-liquid}) is permissible to convert between $C_{V}$ and $C_{P}$ for the liquid Na.
The behavior of $C_{P}$ that initially decreases and then turns to increase as $T$ increases can be explained by the fact that the decrease in $C_{V}$ with $T$ is overcompensated for by the increase in the thermal expansion. It is the sign to approach the critical point.


\section{Simulation results}
\label{sec:Results}

\subsection{Melting}
\label{sec:melting}
The critical issue for the present MD simulations is determining the melting temperature, $T_{m}$, which is by no means a trivial task \cite{Shirai-MeltingT}. Although FPMD simulations do not involve fitting parameters, there are operational parameters that control the quality of the calculations. The obtained results are influenced by these parameters to some extent. It is the authors' experience that the $T_{m}$ value is indeed subjected to these parameters and that FPMD simulations often overestimate the $T_{m}$ value \cite{Shirai-MeltingT}.
We examined the parameter dependence with respect to the cell size, the $k$-mesh for the integration over the Brillouin zone, the cell volume, and the effect of electron occupation. These considerations are described in Supplemental Material.
Finally, we obtained an agreement in $T_{m}$ by the following setup: a cell size of $N=128$, $2^{3}$ $k$-mesh, and an $E_{\rm cut}$ of 18 Ry for the planewave expansion. The authors do not claim that the present setup is good enough to fully describe the liquid state of Na, but mention that the present convergence is at a level sufficient to derive meaningful results. In the following, all the reported calculations were obtained under these conditions.

\begin{figure}[ht!]
\centering
    \includegraphics[width=120mm, bb=0 0 380 500]{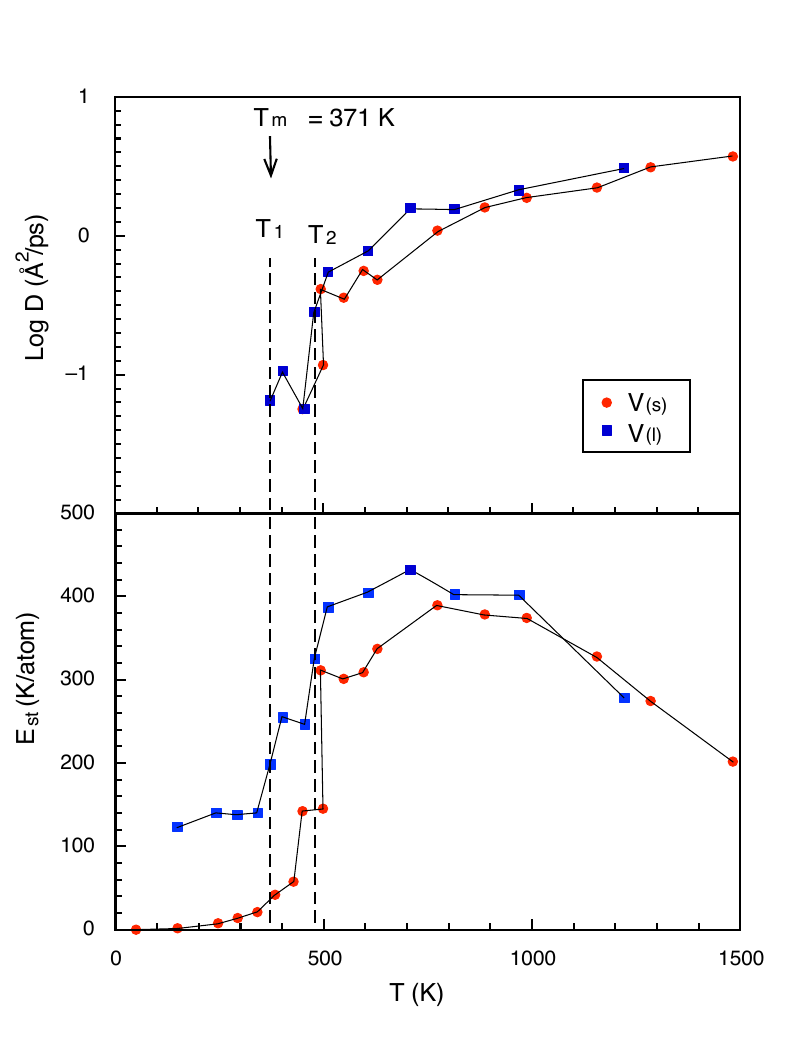} 
\caption{The structural energy, $E_{\rm st}$, and the diffusion constant, $D$, of Na with $N=128$. Two cell volumes, which correspond to those of the solid ($V_{(s)}$, red circles) and liquid ($V_{(l)}$, blue squares), respectively, are examined. In the transition region from $T_{1}$ and $T_{2}$, the system exhibits an unstable behavior.
}
\label{fig:Est-LnD}
\end{figure}

The melting temperature is best determined by a combination of the structural energy, $E_{\rm st}$, and the diffusion constant, $D$ \cite{Shirai-MeltingT}. 
Figure \ref{fig:Est-LnD} shows $E_{\rm st}$ and $D$ as functions of $T$. As can be seen, $E_{\rm st}$ exhibits a significant change at the phase transition. At the temperature where $E_{\rm st}$ increases sharply, a nonzero value $D$ begins to appear.
Two volumes were examined, i.e., the volume of solid, $V_{(s)} \ = 22.69 \ {\rm cm^{3}/mol}$, and that of solid, $V_{(l)} \ =24.84 \ {\rm cm^{3}/mol}$. 
$T_{m}$ is determined by the onset of $D$, when $V_{(l)}$ is employed. Nonzero value $D$ of the liquid appears at $T=371$ K, which just agrees with the experimental value of 371 K within the numerical accuracy.
If $V_{(s)}$ is employed, the obtained value $T_{m}$ turns to be too much overestimation. 

It is noted that, in the temperature region from $T_{1}$ and $T_{2}$ shown in the figure, both $E_{\rm st}$ and $D$ exhibit oscillatory behaviors. This is the common behavior for adiabatic MD simulations with a finite size $N$ \cite{Shirai22-Silica,Shirai22-SH,Shirai-MeltingT}. This region is referred to as {\it the transition region}, where the system becomes thermodynamically unstable. The oscillatory behavior is due to the finite size of cell and thus varies as $N$ changes. The width of the transition region should vanish as $N \rightarrow \infty$. Discussing the detailed $T$ dependence of energies at this region is accordingly not meaningful.

\begin{figure}[ht!]
\centering
    \includegraphics[width=120mm, bb=0 0 400 250]{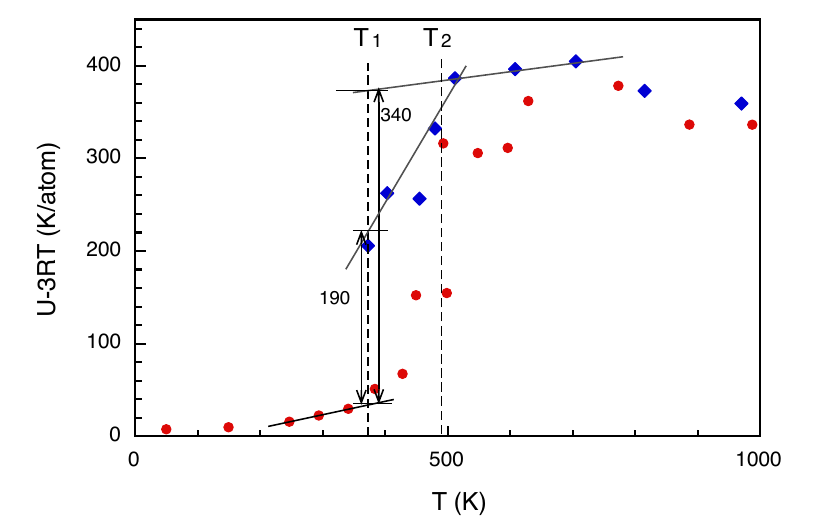} 
\caption{Plot of $U-3RT$ as a function of $T$. The symbols are the same as in Fig.~\ref{fig:Est-LnD}. Two extrapolation methods are indicated with the values obtained for $H_{m}$.
}
\label{fig:U-3RT}
\end{figure}
The melting enthalpy $H_{m}$ is determined by the abrupt increase in $U(T)$ at $T_{m}$. 
To obtain a good resolution, we subtracted an offset $3RT$ from $U(T)$, as shown in Fig.~ \ref{fig:U-3RT}. 
However, the width of the transition region smears this abrupt change, which makes it difficult to estimate $H_{m}$ accurately. Interpolating $U(T)$ within the transition region, we obtain a too small value of $\Delta U=$ 190 K/atom. 
Because the detailed $T$ dependence on $U$ in the transition region is not reliable, an extrapolation from the outside of the transition region ($T>T_{2}$) is required. However, the extrapolation far from $T_{m}$ increases uncertainty. By extrapolating $U(T)$ from the region $500 < T < 700$ K, we obtain $\Delta U=$ 340 K/atom. Note that this also has uncertainty because the region of the linear relation is distant from $T_{m}$. Tentatively, we determine $H_{m}$ according to the average of these two values, namely, $H_{m} = 265$ K/atom, which corresponds to 22.8 meV. This value is to be compared with the experimental value of 26 meV \cite{Alcock94}. In any rate, an improvement in the determination of $T_{m}$ in order not to depend on the way of extrapolation is required for future study. 
From the calculated values of $T_{m}$ and $H_{m}$, the melting entropy, $S_{m}$, is determined as,
\begin{equation}
S_{m} = \frac{H_{m}}{T_{m}} = 0.71 R.
\label{eq:Sm}
\end{equation}
This value is in a good agreement with the experimental value of 0.73 $R$ in \cite{Wallace97a}.

\begin{figure}[ht!]
\centering
    \includegraphics[width=120mm, bb=0 0 360 260]{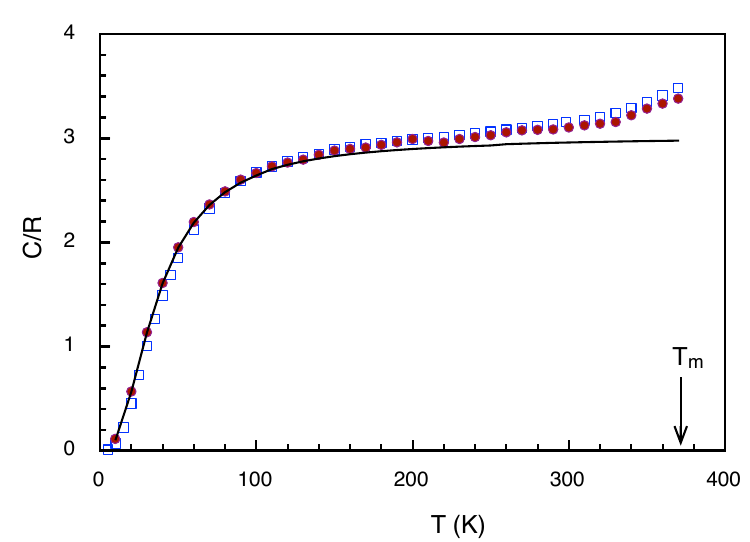} 
\caption{Calculated isochoric specific heat $C_{V}$ of the solid Na (purple circles) and the phonon contribution $C_{\rm ph}$ (solid line). The experimental data (open squares) are those of Fig.~\ref{fig:Alcock}
}
\label{fig:Cv-calc-solid}
\end{figure}

\subsection{Solid state}
\label{sec:solid-result}
Next, the specific heat is calculated by treating the solid and liquid states separately.
For solids, the phonon contribution $C_{\rm ph}$ is easy to calculate. The analytical formula is available as Eq.~(\ref{eq:C-phonon}). The result is plotted in Fig.~\ref{fig:Cv-calc-solid} alongside the experimental data obtained in Fig.~\ref{fig:Alcock}.
At low temperatures, $T<\Theta_{D}$, $C_{\rm ph}$ is virtually the same as $C_{V}$, which indicates that the harmonic approximation is valid. The calculated value is slightly larger than the experimental value at $T<80$ K. This is stemmed from the resolution limit for the acoustic branch of phonon dispersion by a small cell size. The lowest-frequency of TA mode that was reproduced by this size was 20 ${\rm cm}^{-1}$. See the phonon spectrum in Supplemental material.
 At high temperatures, $T>200$ K, $C_{\rm ph}$ is already saturated at the classical limit, $3R$.

\begin{figure}[ht!]
\centering
    \includegraphics[width=100mm, bb=0 0 360 260]{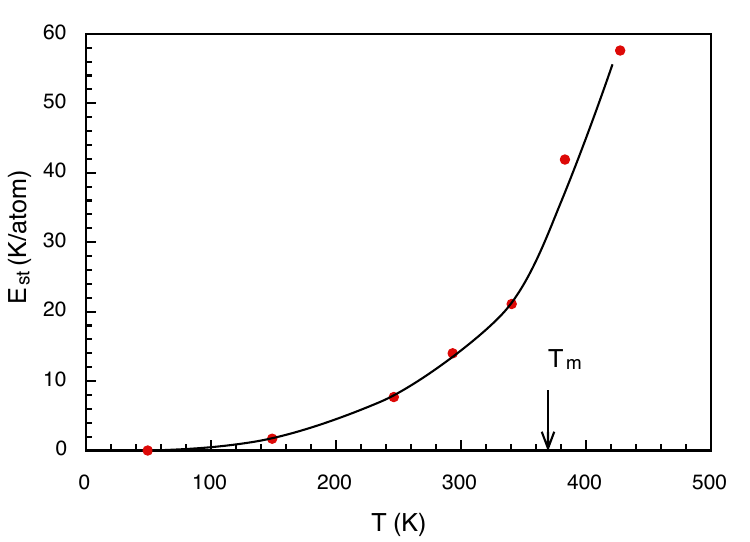} 
\caption{The structural energy, $E_{\rm st}$, of the solid Na. 
}
\label{fig:Est-solid}
\end{figure}
The remaining part $C_{\rm st}$ of $C_{V}$ was obtained by the numerical derivative of $E_{\rm st}$, which is plotted in Fig.~\ref{fig:Est-solid}. As shown, appreciable values of $C_{\rm st}$ appear only at high temperatures, {\it i.e.}, $T>200$ K. 
Adding $C_{\rm st}$ to $C_{\rm ph}$, $C_{V}$ of the solid Na is obtained, which is shown in Fig.~\ref{fig:Cv-calc-solid}. Note that the small fluctuations in the calculated $C_{V}$ are due to the error of the numerical derivative of $E_{\rm st}$. At $T>200$ K, the calculated value is slightly less than the experimental values by about $0.1 R$. The electronic contribution, $C_{\rm el}$, is of this order of magnitude, and hence the agreement would be improved by considering $C_{\rm el}$ into account.

\subsection{Liquid state}
\label{sec:liquid-result}

\begin{figure}[ht!]
\centering
    \includegraphics[width=100mm, bb=0 0 380 260]{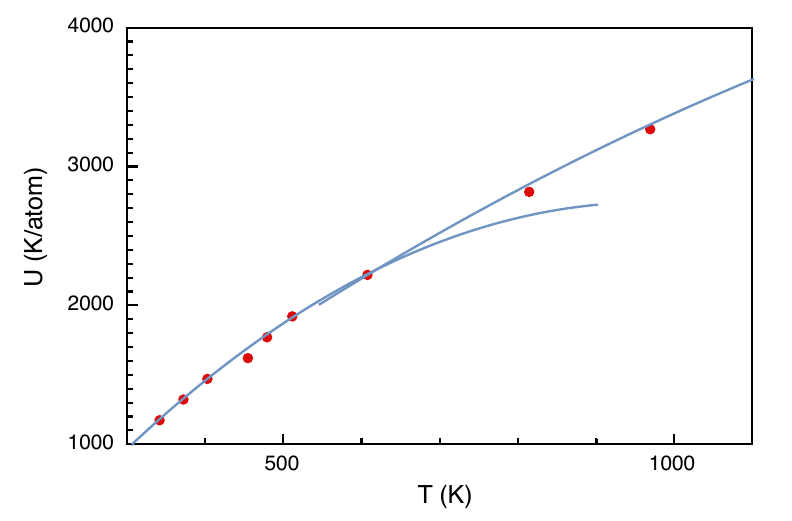} 
\caption{$U(T)$ of the the liquid Na. The data are fitted by quadratic equations in two regions.
}
\label{fig:Uliq-intp}
\end{figure}
For the liquid state, only the total energy is the uniquely determined quantity; thus, the specific heat must be calculated directly from the temperature dependence on $U(T)$. 
The calculated $U(T)$ exhibits fluctuations, which are primarily due to the small size of cell \cite{Shirai-MeltingT}. This is particularly serious in the transition region, where the error in $U(T)$ is magnified in $C_{V}$ by taking the derivative with $T$.
Thus, we adopt a smoothing method to remove the error caused by this fluctuation. Here, the second-order polynomials were used to fit $U(T)$: $U(T)=a_{0} + a_{1}T + a_{2}T^{2}$. The transition region differs from the liquid state region; thus, it was impossible to fit in the entire $T$ range using a single polynomial. Therefore, the fitting was performed for the two regions of $T$, as shown in Fig.~\ref{fig:Uliq-intp}. The values of the coefficients are: for $T<600$ K, $a_{0}=-1000.3$, $a_{1}=7.731$, and $a_{2}=-3.999 \times 10^{-3}$; for $T>600$ K, $a_{0}=175.9$, $a_{1}=3.755$, and $a_{2}=-6.464 \times 10^{-3}$. It is evident that the error is significant in the transition region, and $C_{V}$ has accordingly large errors in this region.

\begin{figure}[ht!]
\centering
    \includegraphics[width=100mm, bb=0 0 380 260]{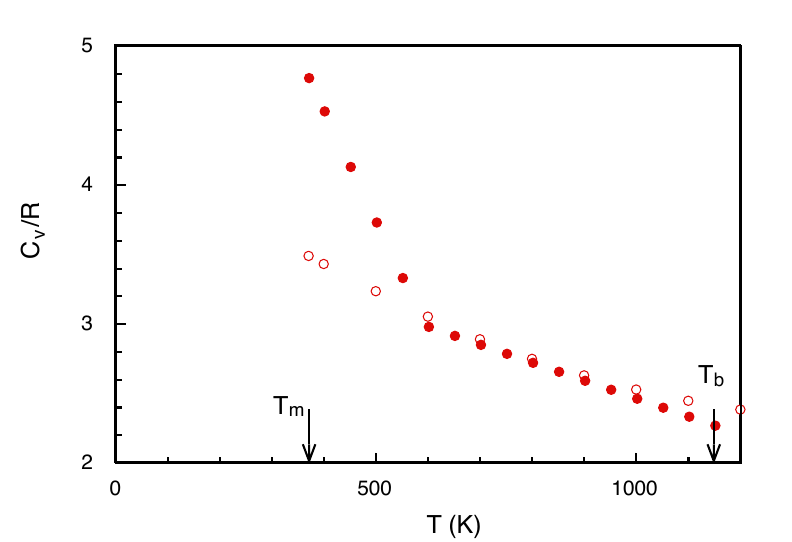} 
\caption{$C_{V}$ of the the liquid Na. The calculated data are indicated by the red circles, and the experimental data in Fig.~\ref{fig:Fink} are indicated by the open circles.
}
\label{fig:Cv-liquid-calc}
\end{figure}

The obtained $C_{V}$ values of the liquid is plotted in Fig.~\ref{fig:Cv-liquid-calc}, alongside the experimental data of Fig.~\ref{fig:Fink}. As expected, the deviation from the experimental value is significant in the transition region. However, for $T>600$K, a good agreement is obtained until close to $T_{b}$. In addition, the decrease in $C_{V}$ with increasing $T$ is well reproduced without introducing any empirical parameters.

\begin{figure}[ht!]
\centering
    \includegraphics[width=120mm, bb=0 0 410 240]{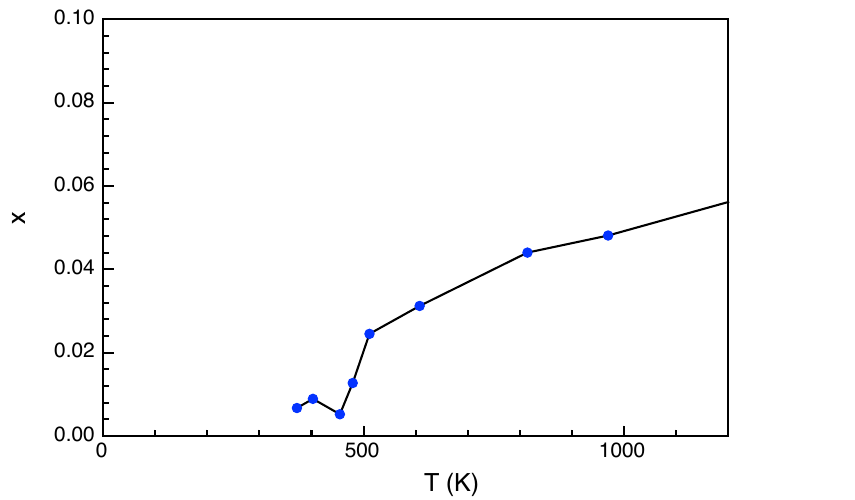} 
\caption{Fraction of zero-frequency modes to the total degree of freedom of atom motions for the liquid Na.
}
\label{fig:x_comp}
\end{figure}

Now, it is a position to discuss whether the phonon model is good for describing this decreasing behavior of $C_{v}$ of the liquid.
When the solid melts, a nonzero component of the zero-frequency modes, $f_{v}(\omega=0)$, appears. The fraction of this component ($x$) to the total degree of the freedom of atom motions ($3N=384$) is shown in Fig.~\ref{fig:x_comp}. As $T$ increases, $x$ increases but is less than 5\% over the entire temperature range. However, we should not interpret this that only 5\% of atoms are moving freely over all of the space inside the material. In reality, all of the atoms are diffusing after melting. This is merely an artifact produced by our definition of the phonon DOS in Eq.~(\ref{eq:phononDOS}). In this definition, even free motions can produce the component, $f_{v}(\omega \neq 0)$, if a collision with the boundary occurs, which turns out to be counted as the frequency component at $\omega \neq 0$. The Fourier analysis does not distinguish scattering from vibrational motions.

\begin{figure}[ht!]
\centering
    \includegraphics[width=120mm, bb=0 0 425 280]{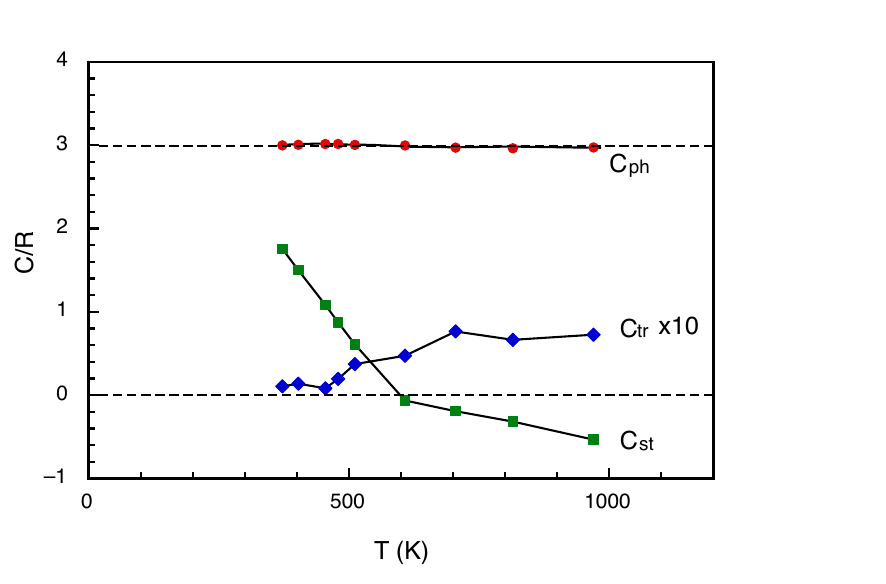} 
\caption{Three components of $C_{V}$ of the liquid Na.
Note that the component $C_{\rm tr}$ is magnified by 10.
}
\label{fig:C_comp_liq}
\end{figure}
The three components, $C_{\rm ph}$, $C_{\rm tr}$, and $C_{\rm st}$ are shown in Fig.~\ref{fig:C_comp_liq}. The component $C_{\rm tr}$ is very small, as expected from the small $x$. Thus, the value of $C_{\rm ph}$ calculated using Eq.~(\ref{eq:phonon-energy-integral}) is virtually constant at the classical limit $3R$. Because of this constancy in $C_{\rm ph}$, $C_{\rm st}$ gets negative values, which corresponds to the decrease in $E_{\rm st}$ with increasing $T$ in the liquid region, as shown in Fig.~\ref{fig:Est-LnD}. Other examples can be found in the literature \cite{Shirai-MeltingT}. Note that the negative specific heat suggests thermodynamic instability \cite{Lynden-Bell99,Landsberg87,Schmidt01,Michaelian07,Michaelian09}.
Obtaining the negative $C_{\rm st}$ in the present case is due to the fact that we overly subtracted from $E_{\rm gs}$ to obtain $E_{\rm st}$ in Eq.~(\ref{eq:Est}). This unphysical result is the consequence that the independent assumption of phonon description is adapted to the liquid state. For liquids, the additive property (\ref{eq:Er=sum-s}) does not hold.

In the phonon model for liquid, it is a common idea that the decrease in $C_{V}$ with $T$ is described by introducing a sort of the cutoff frequency, $\omega_{c}$, below which the corresponding potential does not contribute the specific heat, in a similar manner to the case of pure translational motions. 
In the model by Trachenko {\it et al.},~\cite{Bolmatov12,Trachenko16}, the transverse modes whose frequencies are lower than $\omega_{c}$---which they referred to as Frenkel frequency---are considered the diffusional motions. As $T$ increases, $\omega_{c}$ increases and accordingly the number of transverse modes that contribute $C_{V}$ decreases as
\begin{equation}
C_{V} = R \left[ 3 - \left( \frac{\omega_{c}}{\omega_{D}} \right)^{3} \right],
\label{eq:Trachenko}
\end{equation}
where $\omega_{D}$ is the Debye frequency. In the Wallace's model, a slightly different interpretation is applied \cite{Wallace98}. He considered the upper bound for the phonon amplitudes that two atom vibrations do not overlap: for solids, this bound becomes infinity. Only those parts of phonons whose amplitudes do not exceed this bound can contribute $C_{V}$.  Hence, the correction energy $E_{b}$ in his model has negative values. This bound is presented by a single parameter $\xi$.
In either models, qualitative descriptions capture to some extent the atom dynamics of liquids. However, from the outset, the independent-particle assumption and thereby the application of the Bose-Einstein statistics are not appropriate. The degree of independence among the atom excitations decreases as $T$ increases.

\subsection{Entropy}
\label{sec:lentropy}
\begin{figure}[ht!]
\centering
    \includegraphics[width=120mm, bb=0 0 400 260]{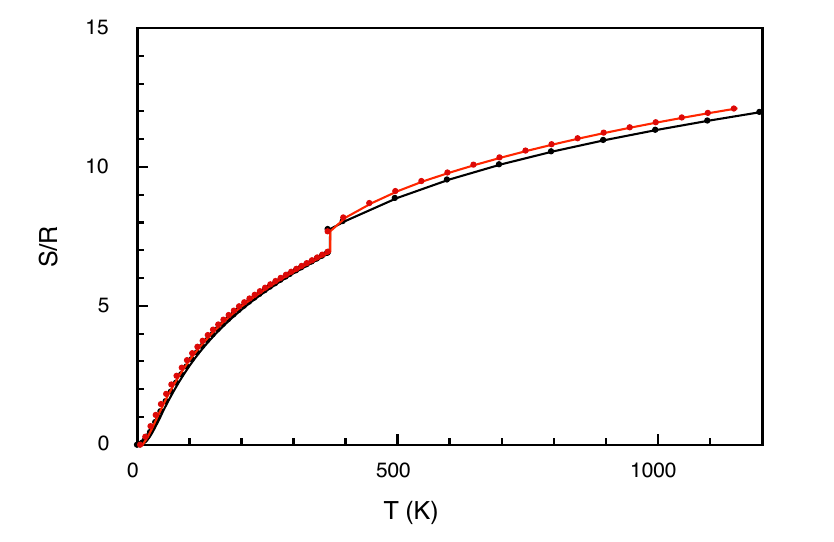} 
\caption{Entropy $S$ of Na over the entire $T$ range of the solid and liquid states. The red dots indicate the calculation, and the black dots indicate the experimental values reported by Alcock {\it et al.}~\cite{Alcock94}.
}
\label{fig:Entropy}
\end{figure}

Everything is in order to calculate the entropy $S$ over all the temperature range. The entropy at 1 atm is obtained by integrating the isobaric specific heat $C_{P}$ at 1 atm according to Eq.~(\ref{eq:thermodynamicS}). For $T<T_{m}$, $C_{P}$ is obtained from the $C_{V}$ value of the solid calculated in Sec.~\ref{sec:solid-result} by adding the experimentally obtained $C_{\rm te}$, as discussed in Sec.~\ref{sec:Experiment}. 
For $T>T_{m}$, we use the experimental relationship given by Eq.~(\ref{eq:Cte-liquid}) for the liquid to convert $C_{V}$ to $C_{P}$.
The result is shown in Fig.~\ref{fig:Entropy}.
For the solid, the agreement is almost satisfactory.
For the liquid, there is a good agreement with respect to the temperature dependence. This could not be obtained if either of the temperature dependence on $C_{V}$ and $C_{\rm te}$ were incorrect. However, there is a slight shift by similar amount over all the $T$ range for the liquid. This is caused by an accumulated error due to the fluctuation in $C_{V}$ at the transition region. In this region, the present calculation overestimates $C_{V}$, as shown in Fig.~\ref{fig:Cv-liquid-calc}.

The enthalpy, $H$, can be obtained directly by integrating $C_{P}$, as $H=\int C_{P} dT$, and then the Gibbs free energy can be obtained by $G=H-TS$. The procedures are trivial, and hence the results are not presented here. In the fundaments of DFT, it is a great challenge to seek explicit expressions for the free energy of liquids by a functional of the atom density profile plus additional variables if necessary \cite{Oxtoby02,Wu-Li07}. The present study gives an alternative method to this approach.

\section{Conclusion}
\label{sec:Conclusion}
The thermodynamic method has been implemented successfully to calculate the entropy of a liquid with a working example of liquid sodium. The central part of the method is the calculation of the specific heat, in which the thermodynamic nature of the liquid is important. In liquids, there are no eigenstates, the fact which invalidates the independent-particle description for the specific heat. The thermodynamic equilibrium is achieved by the dynamic balance between incessant creation and destruction of atom excitations. As a result, atom relaxations control the $U-T$ relation, and the decrease in $C_{V}$ with increasing $T$ must be explained on this basis. This motivates the use of adiabatic simulations. The decrease in $C_{V}$ of the liquid with increasing $T$ is well reproduced by this method. The previously proposed models by phonons lead to an unphysical result of the negative specific heat of the structural energy, and this fallacy occurs because the independent-particle assumption is adapted to a liquid for which no eigenstates are defined.

An advantage of the total-energy approach is that the method is equally applicable to mixed liquids too. The mixing entropy for non-regular liquids is very difficult to calculate using the statistical mechanics method. The total-energy approach combined with the thermodynamic definition of entropy, Eq.~(\ref{eq:thermodynamicS}), provides the total entropy even for mixed liquids, with help of the third law of thermodynamics.
Technically, a problem in the present method is the finite width of the transition region, which impedes the accurate determination of $T_{m}$ and $H_{m}$. In principle, this problem can be solved by increasing the cell size to infinity. $O(N)$ methods may help to reduce the computation costs by increasing the system size \cite{Bowler12,Tanaka24}. Cleaver methods without sacrificing computational cost are desirable.

\section*{Acknowledgment}
The authors thank Prof.~Trachenko (Queen Mary Univ.~London) for discussing the specific heat of glasses and liquids, Prof.~Sugawara (Akita Univ) for the calorimetric methods, Dr. Yamazaki (Osaka Univ.) and Dr. Koga (ASMS Inc.) for the MD simulation.
The authors thank Enago (www.enago.jp) for the English language review.


\begin{thebibliography}{67}
\bibitem[{not()}]{note-effectofbarrier}
\bibinfo{note}{The influence of the energy barriers on the population $P_{i}$
  is demonstrated by extremely simple models of two- or three-level system
  \cite{Peyrard20,Bisquert05}. However, these models are cited here only
  because of the utility indicating the influence of the energy barriers on the
  population $P_{i}$ and therefore should not be considered as models for the
  liquid state.}
\end{thebibliography}

\end{document}